\documentclass[a4paper,11pt]{article}
\usepackage{pos}

\title{Nucleon Spin Sum Rules and Spin Polarizabilities at low $Q^2$}

\author*[a]{A. Deur}

\affiliation[a]{Thomas Jefferson National Accelerator Facility\\
  Newport News, Virginia 23606, USA,}

\emailAdd{deurpam@jlab.org}

\abstract{We report on recently published experimental studies on spin sum rules, namely the generalized Gerasimov-Drell-Hearn, Bjorken, 
Burkhardt-Cottingham, Schwinger, and generalized spin polarizability sum rules. The data were taken at Jefferson Lab in Halls A and B by 
experiments E97-110 and EG4, respectively. They covered the very low $Q^2$ domain, down to $Q^2$$\simeq$0.02~GeV$^2$, where Chiral 
Effective Field Theory ($\chi$EFT) predictions should be valid. While some of the obervables agree with the state-of-the-art $\chi$EFT 
predictions, others are in tensions, including the Longitudinal-Transverse interference polarizability $\delta_{\rm LT}^n(Q^2)$ for which $\chi$EFT 
prediction was expected to be robust. This suggests that $\chi$EFT does not yet consistently describes nucleon spin observables, even in 
the very low $Q^2$ domain covered by the experiments.}

\FullConference{%
 $25^{\rm th}$ International Symposium on Spin Physics \\
 24-29 September 2023 \\
 Duke University, Durham, NC, USA
}

\begin{document}
\maketitle

\section{Spin sum rules}
In Quantum Field Theory, sum rules (SR) are relations between an integral of a dynamical quantity, such as a cross section or structure function, 
and a global property of the object of study, such as its mass or spin. 
SR are used to test the theory and/or the hypotheses from which they are derived. For example, the Bjorken SR~\cite{Bjorken:1966jh} 
has been instrumental in testing the validity of QCD using spin observables~\cite{Gross:2022hyw}. 
Likewise, the Gerasimov-Drell-Hearn (GDH) SR~\cite{Gerasimov:1965et} and related integrals are used to test effective theories  
of the strong force, while the Ellis-Jaffe SR~\cite{Ellis:1973kp} has been used to test the hypothesis that strange quarks 
contributed negligibly to the nucleon spin structure~\cite{Deur:2018roz}.
Another important use of SR is to provide a means to measure the global property involved in the SR, e.g., a generalized spin polarizability, 
for which no direct measurement method is known.

Here, we will discuss spin SR, where the integral is over spin structure function(s). Specifically, we report on experimental results from 
Jefferson Lab (JLab) experiments E97-110~\cite{Sulkosky:2019zmn} and EG4~\cite{Adhikari:2017wox, Zheng:2021yrn, Deur:2021klh} on the 
Bjorken~\cite{Bjorken:1966jh}, Burkhardt-Cottingham (BC)~\cite{Burkhardt:1970ti}, Schwinger~\cite{Schwinger:1975ti}, 
and generalized spin polarizability SR~\cite{Guichon:1995pu}, as well as the GDH SR and its generalizations~\cite{Gerasimov:1965et, Anselmino:1988hn}.
These SR involve either the first moment (Bjorken, BC, Schwinger, and GDH SR) or higher moments (generalized spin polarizability SR) of the 
spin structure functions $g_1(x,Q^2)$ and/or $g_2(x,Q^2)$, or equivalently, the inclusive partial polarized cross-sections $\sigma_{\rm TT}(\nu,Q^2)$ 
and/or $\sigma_{\rm LT}'(\nu,Q^2)$~\cite{Deur:2018roz}. (Here, $Q^2$ is the squared 4-momentum exchanged in inclusive lepton scattering, 
$\nu$ is the corresponding exchanged energy, and $x$ is the Bjorken scaling variable.)
Experiments E97-110 and EG4 precisely measured the $Q^2$-behavior of the SR moments at very low $Q^2$ values, 
with the express aim of testing Chiral Effective Field  Theory ($\chi$EFT)~\cite{Bernard:1995dp} with spin-dependent observables.

 \section{Chiral Effective Field Theory}
 $\chi$EFT is the leading effective theory of the strong force at long distances, specifically where the most relevant degrees of freedom are the 
 hadronic ones, rather than the partonic ones. Consequently, $\chi$EFT concerns itself with the initial level of complexity emerging from the Standard 
 Model. As such,  $\chi$EFT is a crucial piece of our global understanding of Nature. This warrants rigorously testing it to ascertain its validity and limitations.
 $\chi$EFT successfully passed these tests for many hadronic and nuclear phenomena~\cite{Bernard:1995dp}. 
 However, late 1990s results from JLab experiments~\cite{Amarian:2002ar, Yun:2002td, Dharmawardane:2006zd, Fersch:2017qrq}
  compared to the $\chi$EFT predictions available at the time~\cite{Bernard:1992nz, Ji:1999pd, Bernard:2002bs, Kao:2002cp} 
 suggested either of issues in the $\chi$EFT description of the nucleon spin observables, or/and that the $Q^2$ range of validity of 
 $\chi$EFT, expected to be at least a few tenths of GeV, was considerably smaller than this for spin observables.
 \begin{table}[ht]
 \center
{\small
\caption{
Late 1990s/early 2000s comparison between moment data and $\chi$EFT predictions. Bold symbols, e.g., $\pmb{\delta_{\rm LT}^n}$, 
signal moments for which $\chi$EFT predictions were expected to be most robust. ``{\color{blue}{\bf{A}}}'' indicates that data and predictions 
agree up to at least $Q^2\simeq0.1$ GeV$^2$, ``{\color{red}{\bf{X}}}'' indicates that they disagree and ``-'' that no calculation was available at the time.
The  $p$+$n$  superscript denotes either deuteron data without break-up channel, 
or proton+neutron data added together with neutron information either from $^3$He or D.
\label{xpt-comp-2002}}
\begin{tabular}{|c|c|c|c|c|c|c|c|c|c|}
\hline 
Ref. & $\Gamma_1^p$ & $\Gamma_1^n$ & $\pmb{\Gamma_1^{p-n}}$ & $\Gamma_1^{p+n}$ &  $\gamma_0^p$ & $\gamma_0^n$ & $\pmb{\gamma_0^{p-n}}$ & $\gamma_0^{p+n}$ & $\pmb{\delta_{\rm LT}^n}$   \tabularnewline
\hline
\hline 
Ji 1999 {\it et al.} \cite{Ji:1999pd}  & {\color{red}\bf{X}} & {\color{red}\bf{X}} & {\color{blue}\bf{A}} & {\color{red}\bf{X}} & - & - & - & - & -  \tabularnewline
\hline 
Bernard {\it et al.} 2002 \cite{Bernard:2002bs} & {\color{red}\bf{X}} & {\color{red}\bf{X}} & {\color{blue}\bf{A}} & {\color{red}\bf{X}} & {\color{red}\bf{X}} & {\color{blue}\bf{A}} & {\color{red}\bf{X}} & {\color{red}\bf{X}} & {\color{red}\bf{X}}\tabularnewline
\hline 
Kao {\it et al.} 2002 \cite{Kao:2002cp}  & - & - & - & - & {\color{red}\bf{X}} & {\color{blue}\bf{A}} & {\color{red}\bf{X}} & {\color{red}\bf{X}} & {\color{red}\bf{X}} \tabularnewline
\hline 
\end{tabular}
}
\vspace{-0.6cm}
\end{table}
Table~\ref{xpt-comp-2002} presents the comparison between results obtained from late 1990s/early 2000s JLab experiments, specifically Hall A E94010~\cite{Amarian:2002ar}, CLAS EG1a~\cite{Yun:2002td}, and EG1b~\cite{Dharmawardane:2006zd, Fersch:2017qrq}.
Here, $\Gamma_1 \equiv \int_0^{1^-} g_1 dx$, viz a generalized GDH sum or, in the particular case of $\Gamma_1^{p-n}$ the Bjorken sum. $\gamma_0$ is the forward spin polarizability and $\delta_{\rm LT}$ is the  generalized LT-interference polarizability.
Clearly, $\chi$EFT predictions were in tension with the 1990s-2000s spin observable data more often than not.
This finding  was particularly puzzling for $\delta_{\rm LT}$. It was expected to allow for robust measurements due to the negligible   
unmeasured low-$x$ contributions for higher moments, and for robust $\chi$EFT predictions because the nucleon resonance $\Delta$ 
contribution (not included in the calculations~\cite{Bernard:1992nz, Ji:1999pd, Bernard:2002bs, Kao:2002cp} due to its complexity) was expected to 
be suppressed in $\delta_{\rm LT}$. Indeed, a similar suppression was also expected for the Bjorken sum $\Gamma_1^{p-n}$~\cite{Burkert:2000qm}, 
which was confirmed by the data~\cite{Deur:2004ti}, as shown in Table~\ref{xpt-comp-2002}.
This led to the question of whether this $\delta_{\rm LT}$ puzzle --and more generally the less-than-ideal state of affaires summarized in 
Table~\ref{xpt-comp-2002}-- was a problem in the $\chi$EFT calculations or if the experiments had not reach yet
the applicability domain of $\chi$EFT, i.e., low enough $Q^2$. To address this question, refined $\chi$EFT calculations were conducted, 
incorporating improved expansion schemes and including the $\Delta$ contribution~\cite{Bernard:2012hb, Lensky:2014dda}. Complementarily on 
the experimental front, a new JLab program was proposed and executed, aiming to reach deeper into the $\chi$EFT  domain and with 
enhance precision.
The outcome from this experimental program is the subject of this proceeding. Given the many observables that were measured, 
we will focus here on just a few representative results. An overview of the program within the larger context of spin structure studies 
at JLab was given in this conference by J.-P. Chen~\cite{JPChen-spin2023} 
and specific results on $^3$He were shown by C. Peng~\cite{E97-He3 data-spin-conf}.
 
 \section{The JLab experimental program on spin sum rules at  low $Q^2$  }

\subsection{The experiments}

The low $Q^2$ spin SR experimental program, which ran at JLab between 2004 and 2012, viz during its 6 GeV era, comprises four inclusive  
doubly-polarized electron scattering experiments, 
two in Hall A and two in Hall B. In Hall A, E97-110 focused on neutron spin using a longitudinally and transversally polarized $^3$He 
target~\cite{Sulkosky:2019zmn}, and E08-027 studied the proton's transverse spin structure with a longitudinally and transversely polarized 
NH$_3$ target~\cite{JeffersonLabHallAg2p:2022qap}. The Hall B experiments, grouped under run group EG4, are E03-006 (E06-017), 
focusing on the proton (neutron and deuteron) longitudinal spin structure with a  longitudinally polarized NH$_3$ (ND$_3$)
target~\cite{Zheng:2021yrn} (\cite{Adhikari:2017wox}). 
These proceedings cover E97-110 and EG4.

To access the very low $Q^2$ domain suitable for $\chi$EFT testing while maintaining the large range of $\nu$ (or equivalently $x$) values necessary 
for forming moments, demands to detect the scattered electrons at forward angles, typically below $10^\circ$. For E97-110, a new magnet (the 
``septum magnet'') was inserted between the target and the Hall A High Resolution Spectrometer, enabling to reach angles to around $5^\circ$ 
compared to the minimum $12.5^\circ$ in standard configuration. For EG4, the target's relocation a meter farther from its standard position 
allowed to reach approximately $4^\circ$. However, at these angles, the detection efficiency is too low for accurate cross-section 
measurements. Therefore, a purpose-built Cerenkov detector was installed in the sector 6
of CLAS, allowing for high-efficiency detection down to about $6^\circ$.

\subsection{Experimental results}
The experiments measured  doubly polarized inclusive inelastic reactions, extracting from them the spin structure functions or 
partial cross-sections. The longitudinal spin structure function $g_1(x,Q^2)$ and partial cross-section $\sigma_{\rm TT}$ 
were obtained in the $0.01<Q^2<0.085$~GeV$^2$ and $0.006<x<0.71$ ranges for both 
the proton~\cite{Zheng:2021yrn}, neutron~\cite{EG4-long paper}, deuteron~\cite{Adhikari:2017wox} and $^3$He~\cite{Sulkosky:2019zmn, E97-He3 data}
Elastic cross-section ($x=1$) was measured as well for some of the $Q^2$ values. The transverse spin structure function $g_2(x,Q^2)$ and 
partial cross-section $\sigma_{\rm LT'}$ were obtain in similar range for the $^3$He~\cite{Sulkosky:2019zmn, E97-He3 data}.

\begin{figure}[ht]
  \centering
    \includegraphics[width=1\textwidth]{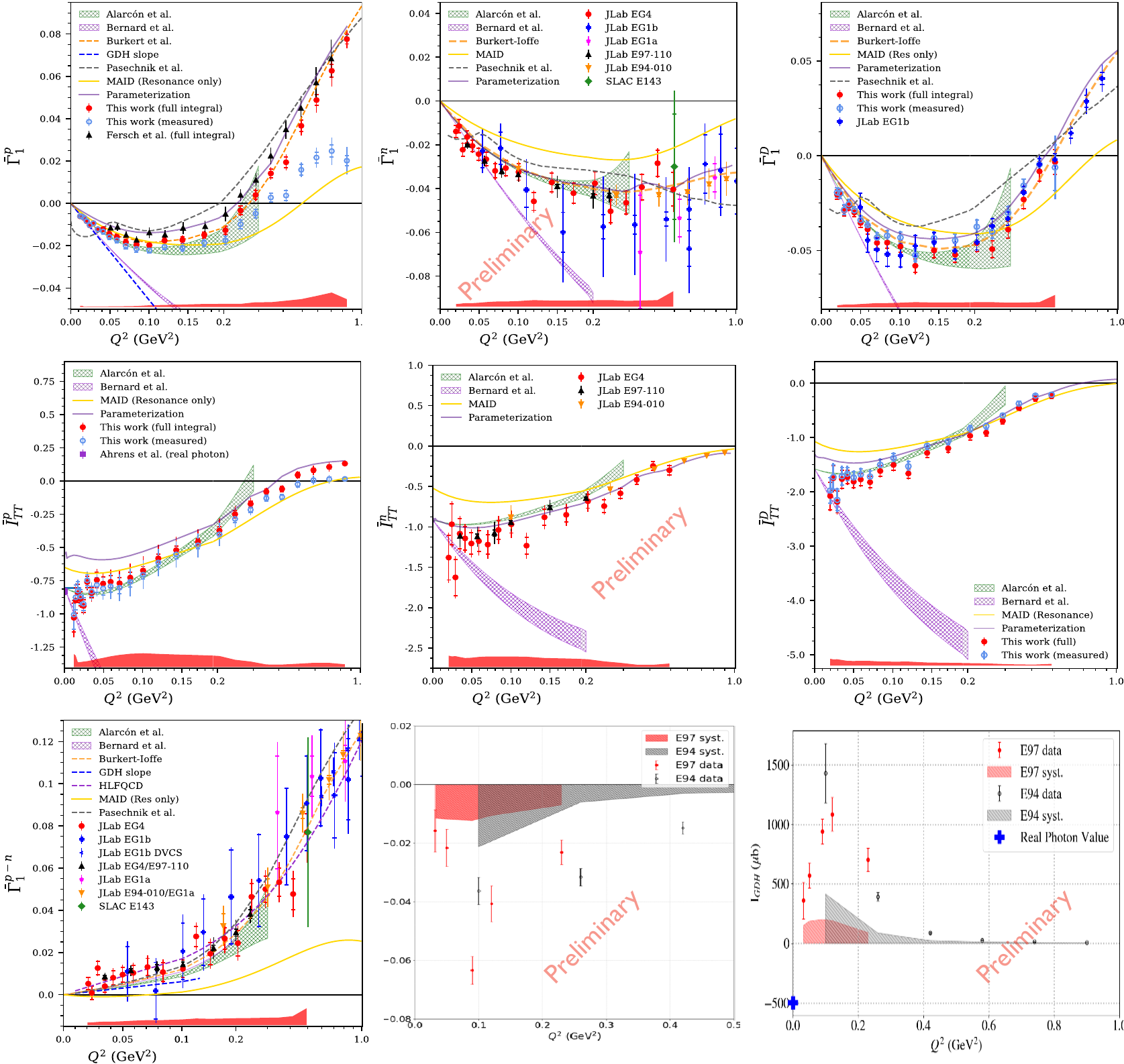}
 \caption{\label{fig:moments-1} 
$Q^2$-dependence of the first moments of the longitudinal spin structure function $g_1$ or of the partial cross-section $\sigma_{\rm TT}$ for proton, 
neutron, D (without break-up channel contribution) and $^3$He. Neutron moments from EG4, and $^3$He ones from E97-110 are preliminary.
First row: the generalized GDH sum $\Gamma_1^p$, $\Gamma_1^n$ and $\Gamma_1^D$.
Middle row: the generalized GDH sum $I_{\rm TT}^p$, $I_{\rm TT}^n$ and $I_{\rm TT}^D$.
Bottom row, left: the Bjorken sum $\Gamma_1^{p-n}$, center and right, generalized GDH sums $\Gamma_1^{^3\rm He}$ and $I_{\rm TT}^{^3\rm He}$. 
The bar over symbols indicates that the elastic contribution is not included in the moments.}
\end{figure}

\begin{figure}[ht]
  \centering
    \includegraphics[width=1\textwidth]{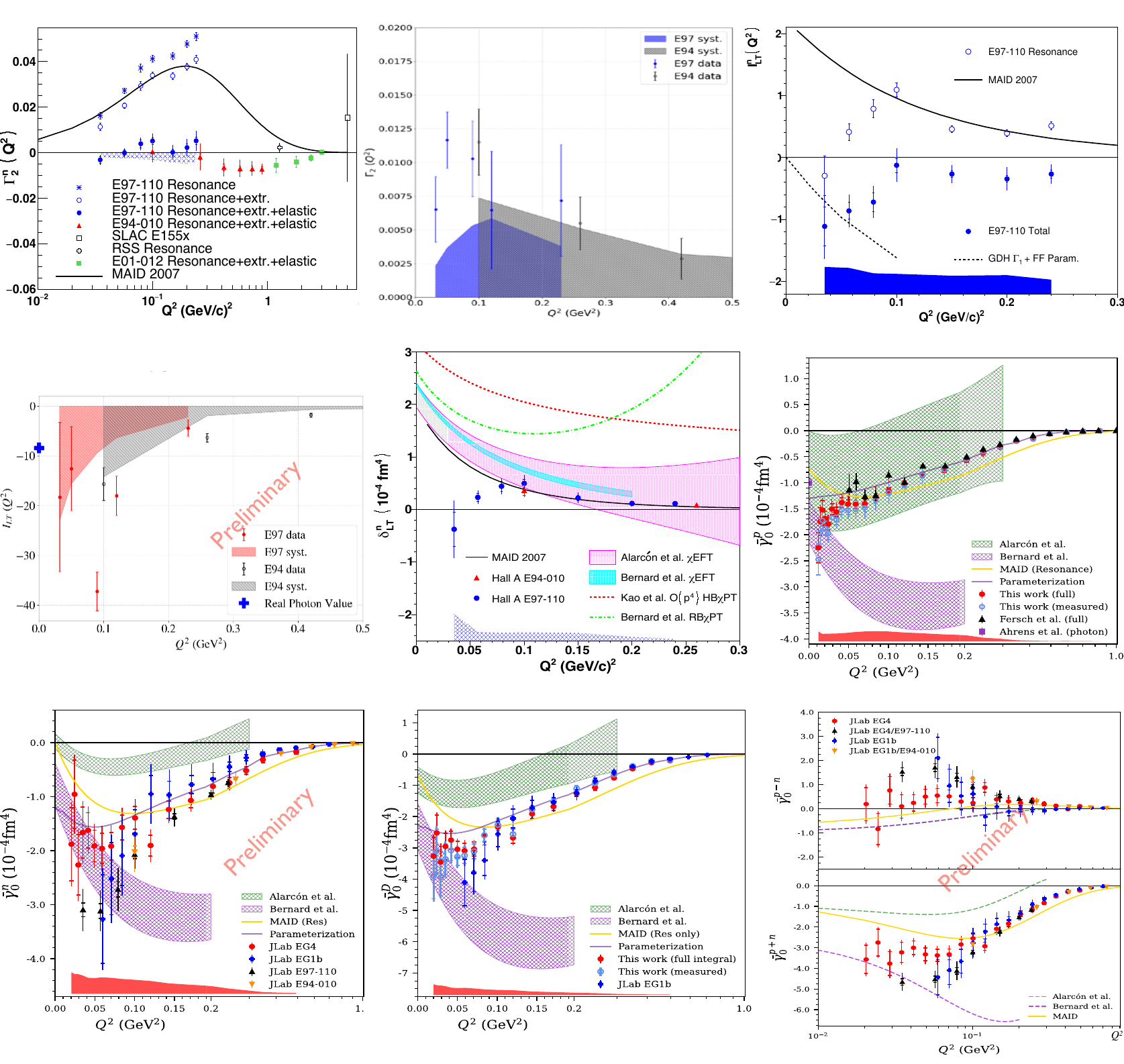}
 \caption{\label{fig:moments-2} 
$Q^2$-dependence of moments for the neutron and $^3$He spin structures. 
Top row, left and middle: first moments of the transverse spin structure function $g_2$: $\Gamma_2^n$ and $\Gamma_2^{^3\rm He}$ (BC SR).
Top row, right: first moment of partial cross-section $\sigma_{\rm LT}$ for neutron: $I^n_{\rm LT}$ (Schwinger SR).
Middle row, left: $I^{^3\rm He}_{\rm LT}$. Central panel: neutron's generalized LT-interference polarizability $\delta^n_{\rm LT}$.
Middle row, right: proton's generalized forward spin polarizability $\gamma^p_0$.
Bottom row, from left to right: $\gamma^n_0$, $\gamma^D_0$ and isospin decomposition of $\gamma_0$.
The bar over symbols indicates that the elastic contribution is not included.
(Neutron moments from EG4, and $^3$He ones from E97-110 are preliminary.)}
\end{figure}

From these quantities, moments for spin SR could be formed. Since it is not experimentally feasible to reach $x$=0, 
the lowest $x$ part of the moments, typically $x$$<$5$\times$$10^{-3}$ for EG4 and E97-110,
was estimated using models~\cite{Fersch:2017qrq, Bass:1997fh}. This contribution is noticeable
only for first moments, for which it remains small in the cases of E97-110 and EG4~(see Figs.~\ref{fig:moments-1} and~\ref{fig:moments-2}). 
It is negligible for higher moments but those are
more sensitive to possible high-$x$ contamination from elastic and quasi-elastic reactions. Efforts were made in E97-110 and EG4 to suppress
radiative tails from these reactions
by reducing as much as possible material thickness around the targets. These tails were carefully modeled and 
subtracted. In addition, for EG4 where material thickness was larger and which spectrometer did not have the high resolution of that of Hall A,  
the EG1b parameterization~\cite{Fersch:2017qrq} was used instead of the data for the high-$x$ contribution to moments, 
specifically in the missing mass range from threshold production to $W=$1.15 GeV. 
The partial moments and the full ones after being complemented for low-$x$ (and large-$x$ for EG4) are shown in Figs.~\ref{fig:moments-1} and~\ref{fig:moments-2}.

Results from EG4 on the generalized GDH sum on the proton, $\Gamma_1^p(Q^2)$, are shown in Fig.~\ref{fig:moments-1}, 
top left panel. Comparing the measured partial moment (open blue circles) to the full one (filled red circles) 
shows that the unmeasured low-$x$ contribution is small. Comparison to the previous experiment, EG1b~\cite{Fersch:2017qrq}, 
shows that the low $Q^2$ reach was decreased by about a factor of 4, and that the data precision is much improved. (The slight systematic 
difference between EG4 and EG1b above $Q^2$$\simeq$0.1~GeV$^2$ is attributed to improved radiative tail subtraction done for 
EG4.) These features make EG4 to provide a clean test of $\chi$EFT. Data and $\chi$EFT predictions agree up to $Q^2$$\simeq$0.04~GeV$^2$ 
(Bernard \textit{et al.}~\cite{Bernard:2012hb}) or $Q^2$$\simeq$0.2~GeV$^2$ (Alarc\'on \textit{et al.}~\cite{Lensky:2014dda}). 
The Burkert-Ioffe~\cite{Burkert:1992tg} and MAID~\cite{Drechsel:1998hk} phenomenological models agree well with data, while that of 
Pasechnik \textit{et al.}~\cite{Pasechnik:2010fg} and the parameterization based on EG1b data~\cite{Fersch:2017qrq} (purple line) agree less well.

Neutron results from E97-110 and EG4 on $\Gamma_1^n(Q^2)$ are shown in Fig.~\ref{fig:moments-1}, 
top middle panel. The new data reach a lowest $Q^2$ smaller by about a factor of 4 for EG4 and of 2 for E97-110 compared to 
previous data, and with a much-improved precision, especially for E97-110. EG4 and E97-110, 
despite  their neutron information stemming  from D and $^3$He respectively, agree well with each other. They also agree with the older data 
(EG1b~\cite{Dharmawardane:2006zd, Fersch:2017qrq}, E94-010~\cite{Amarian:2002ar}). The new data agree with $\chi$EFT up to 
$Q^2$$\simeq$0.06~GeV$^2$ for prediction~\cite{Bernard:2012hb} and $Q^2$$>$0.4~GeV$^2$ for that of~\cite{Lensky:2014dda}. For the neutron, the 
Burkert-Ioffe phenomenological model and the parameterization based on EG1b agree well with the data, 
while the MAID and the Pasechnik \textit{et al.} models match them less well.

The isovector part of $\Gamma_1$, viz, $\Gamma_1^p-\Gamma_1^n \equiv \Gamma_1^{p-n}$, provides the moment of the Bjorken sum 
rule~\cite{Bjorken:1966jh}. The combined E97-110 and EG4 data~\cite{Deur:2021klh} are shown in Fig.~\ref{fig:moments-1}, bottom left 
panel. The new data agree well with the previous ones from E94010, EG1a, EG1b, EG1dvcs~\cite{Deur:2004ti} and SLAC E143~\cite{Abe:1998wq}. 
They also agree with $\chi$EFT, supporting the expectation that its predictions are robust because the $\Delta$ 
cancel in $\Gamma_1^{p-n}$~\cite{Burkert:2000qm}. Yet, these predictions and the available models tend to be below the data for $Q^2$$<$0.1~GeV$^2$, 
except for the HLFQCD model~\cite{Brodsky:2010ur}. The Bjorken sum data have been used to extract the QCD effective charge in the 
non-perturbative domain~\cite{Deur:2005cf}, agreeing well with theoretical expectations~\cite{Deur:2016tte, Gross:2022hyw}.

Another possible GDH generalization is $I_{\rm TT}(Q^2) \equiv \frac{2M^2}{Q^2}\int_0^{x_0}\Big[g_1-\frac{4M^2}{Q^2}x^2g_2\Big]dx$,
with $M$ the target particle mass. 
Compared to the other generalization $\Gamma_1(Q^2)$, there is no suppressing $Q^2$ factor for $I_{\rm TT}$, but it
involves the transverse spin structure function $g_2$, which measurement requires a transversely polarized target and was not measured in EG4. 
Yet, $g_2$ enters at the second order, allowing us to model it without making the measurement of $I_{\rm TT}$ too model-dependent. 
(For $I_{\rm TT}^n(Q^2)$, which was not presented at Spin 2023 for lack of time, E97-110 measured $g_2$ so there is no associated
model-dependence in that case.)
Results on $I_{\rm TT}^p$ from EG4 are shown in Fig.~\ref{fig:moments-1}, middle row, left panel. Similar conclusions as with $\Gamma_1^p$ are 
reached when comparing EG4's $I_{\rm TT}^p$ data to $\chi$EFT. On the other hand, none of the available models (MAID and the EG1b 
parameterization) agree well with the data. Extrapolating the lowest $Q^2$ points of $I_{\rm TT}^p$ to $Q^2$=0 provides an independent check of the 
GDH SR validity, using a different method (inclusive data) than photoproduction experiments (exclusive data)~\cite{Dutz:2004zz}. The EG4 
extrapolation, $I_{\rm TT}^{p~\rm{EG4}}(0)$=-0.798$\pm$0.042, agrees with the GDH SR, $I_{\rm TT}^{p~\rm{theo}}(0)$=-0.804$\pm$0.000, 
testing the SR with a precision similar to the photoproduction method, 
$I_{\rm TT}^{p~\rm{MAMI}}(0)=$-0.832$\pm$0.023(stat)$\pm$0.063(syst)~\cite{Dutz:2004zz}.

We turn now to higher moments. The generalized forward spin polarizability 
$\gamma_0(Q^2) = \frac{16\alpha M^2}{Q^6}\int_0^{1^-} x^2\big[g_1-\frac{4M^2}{Q^2}x^2 g_2\big]dx$ was measured both on the proton by 
EG4~\cite{Zheng:2021yrn} (bottom row, left panel in Fig.~\ref{fig:moments-2}) and on the neutron by E97-110 (from $^3$He)~\cite{Sulkosky:2019zmn} 
and EG4 (from D)~\cite{EG4-long paper} (bottom row, middle panel).
Focusing on $\gamma_0^n$, we observe that while the E97-110 and EG4 data agree well with the previous data at larger $Q^2$ from 
E94-010~\cite{Amarian:2002ar} and EG1b~\cite{Dharmawardane:2006zd}, they only marginally agree with each other in the newly covered low 
$Q^2$ range. The comparison is performed with EG4 systematic uncertainties added in quadrature, but when linearly added, the tension 
vanishes. The $\chi$EFT result of~\cite{Lensky:2014dda} now disagrees with the data, in contrast with first moment results, while those 
of~\cite{Bernard:2012hb} continue to agree for the lowest $Q^2$ points, as observed with first moments. The EG1b parameterization matches 
the new data, but MAID starkly disagrees.

Measuring the other generalized spin polarizability, $\delta_{\rm LT}(Q^2) = \frac{16\alpha M^2}{Q^6}\int_0^{1^-} x^2\left[g_1 + g_2\right]dx$, 
requires a transversely polarized target, so it was measured only by E97-110 (and E08-027, see~\cite{JPChen-spin2023} 
for the Spin-2023 report). The new data on $\delta_{\rm LT}^n$ are shown in Fig.~\ref{fig:moments-2}, 
center panel. They agree well with the previous data (E94-010~\cite{Amarian:2002ar}) at larger $Q^2$ and with the latest 
$\chi$EFT predictions~\cite{Lensky:2014dda, Bernard:2012hb} and MAID in that region. However, these predictions disagree with the data in the 
newly covered low $Q^2$ range. Interestingly, the data on the corresponding first moment 
$I^n_{\rm LT}(Q^2) = \frac{8 M^2}{Q^2}\int_0^{1^-} \left[g_1 + g_2\right]dx$ agrees with the Schwinger SR, 
$I^n_{\rm LT}(Q^2\to0) \to 0$, see Fig.~\ref{fig:moments-2}, top right panel. Therefore, the 
"$\delta_{\rm LT}$ puzzle" that instigated the low-$Q$ program remains.

\subsection{New tests of $\chi$EFT}
Table~\ref{xpt-comp-2002} can now be updated with refined $\chi$EFT calculations~\cite{Bernard:2012hb, Lensky:2014dda} and data from 
the JLab low-$Q^2$ program~\cite{Sulkosky:2019zmn, Adhikari:2017wox, Zheng:2021yrn, Deur:2021klh}. 
Comparing Tables~\ref{xpt-comp-2002} and~\ref{xpt-comp-2022} reveals a clear improvement since the early 2000s 
(again, the statement whether $\chi$EFT and data agree or not refers to the range $Q^2$$\lesssim$0.1~GeV$^2$). Yet, there are still many 
instances of disagreements despite the improved calculations and despite the new data now being well into $\chi$EFT's expected validity domain.
\begin{table}[ht]
 \center
{\small
\caption{Same as table \ref{xpt-comp-2002} but now including the  latest $\chi$EFT predictions~\cite{Bernard:2012hb, Lensky:2014dda} and data~\cite{Sulkosky:2019zmn, Adhikari:2017wox, Zheng:2021yrn, Deur:2021klh}.
\label{xpt-comp-2022}}
\begin{tabular}{|c|c|c|c|c|c|c|c|c|c|}
\hline 
Ref. & $\Gamma_1^p$ & $\Gamma_1^n$ & $\pmb{\Gamma_1^{p-n}}$ & $\Gamma_1^{p+n}$ &  $\gamma_0^p$ & $\gamma_0^n$ & $\pmb{\gamma_0^{p-n}}$ & $\gamma_0^{p+n}$ & $\pmb{\delta_{\rm LT}^n}$   \tabularnewline
\hline
\hline 
Ji 1999 {\it et al.} \cite{Ji:1999pd}  & {\color{red}\bf{X}} & {\color{red}\bf{X}} & {\color{blue}\bf{A}} & {\color{red}\bf{X}} & - & - & - & - & -  \tabularnewline
\hline 
Bernard {\it et al.} 2002 \cite{Bernard:2002bs} & {\color{red}\bf{X}} & {\color{red}\bf{X}} & {\color{blue}\bf{A}} & {\color{red}\bf{X}} & {\color{red}\bf{X}} & {\color{blue}\bf{A}} & {\color{red}\bf{X}} & {\color{red}\bf{X}} & {\color{red}\bf{X}} \tabularnewline
\hline 
Kao {\it et al.} 2002 \cite{Kao:2002cp}  & - & - & - & - & {\color{red}\bf{X}} & {\color{blue}\bf{A}} & {\color{red}\bf{X}} & {\color{red}\bf{X}} & {\color{red}\bf{X}} \tabularnewline
\hline 
Bernard {\it et al.} 2012 \cite{Bernard:2012hb}  & {\color{red}\bf{X}} & {\color{red}\bf{X}} & {\color{blue}\bf{A}} & {\color{red}\bf{X}} & {\color{red}\bf{X}} & {\color{blue}\bf{A}} & {\color{red}\bf{X}} & {\color{red}\bf{X}} & -\tabularnewline
\hline 
Alarc\'on {\it et al.} 2020 \cite{Lensky:2014dda} & {\color{red}\bf{X}} & {\color{blue}\bf{A}} &  {\color{blue}\bf{A}} & {\color{blue}\bf{A}} &  {\color{blue}\bf{A}} & {\color{red}\bf{X}} & {\color{red}\bf{X}} & {\color{red}\bf{X}} &  {\color{blue}\bf{$\sim$ A}} \tabularnewline
\hline
\end{tabular}
}
\vspace{-0.6cm}
\end{table}

\section{Summary and conclusion}
The JLab experiments E97-110 and EG4 have provided high-precision nucleon spin structure data in the very low $Q^2$ domain, 
where $\chi$EFT is expected to describe well the strong force. In general, the new data agree well with those of previous experiments that covered larger 
$Q^2$ values. E97-110 and EG4 also agree with each other (albeit marginally for $\gamma_0^n$ for 
the lowest $Q^2$ points).
When used to test spin SR, namely the GDH, BC and Schwinger SR, the new data agree within uncertainties with the 
SR expectations. On the other hand, there is mixed agreement/disagreement with the latest $\chi$EFT predictions, depending 
on which observable, $Q^2$ range, and calculations are considered. In particular, the ``$\delta_{\rm LT}$ puzzle'' remains, and 
$\gamma_0^{p-n}$ disagrees too with $\chi$EFT.
One reason for this could be that maybe the data are less accurate than asserted. In fact, low $Q^2$ moment 
measurements are hard: they are necessarily very forward-angle experiments, where backgrounds are typically large; 
they demand extensive $x$ coverage and are affected by the impossibility to measure down to $x$=0, like any experiment 
measuring moments (albeit generally to a lesser extent than larger $Q^2$ measurements); they are subject to high-$x$ 
contamination by radiative tails that demand careful subtraction, etc.
However, the experiments constituting the low-$Q$ JLab program were independently ran, employing very different detectors and methods 
of analysis. Furthermore, they measured observables with variable sensitivity to the low-$x$ or high-$x$ issues. Yet, their experimental 
message is consistent. 
From this and the fact that some of the $\chi$EFT predictions disagree with 
each other, we may conclude that although $\chi$EFT is successful in many instances, it is challenged by polarized data.

This is hindering our endeavor toward a complete description of Nature at all levels because $\chi$EFT is the leading approach to manage the 
first level of complexity arising from the Standard Model, viz. the strong force described with hadronic degrees of 
freedom. This is a serious issue since it would be, e.g., akin to atomic physics not providing the theoretical foundations of chemistry.
$\chi$EFT calculations are not easy. Going to next order of the chiral perturbation series to see whether the current issue comes 
from slow convergence will be very challenging. It would help with this problem if we could get low $Q^2$ predictions 
for spin observables from other non-perturbative 
approaches to QCD, such as Lattice QCD, the Dyson-Schwinger equations~\cite{Maris:2003vk}, or AdS/QCD~\cite{Brodsky:2014yha}.

\begin{acknowledgments}
This material is based  upon work supported by the U.S. Department of Energy, Office of Science, 
Office of Nuclear Physics under contracts DE-AC05-06OR23177. 
\end{acknowledgments}

\end{document}